\documentclass[aps,prd,floatfix,superscriptaddress,showpacs,notitlepage]{revtex4-1}

\usepackage{graphicx}
\usepackage{dcolumn}
\usepackage{bm}
\usepackage{epsfig}

\newcommand\eqn[1]{Eq.\,(\ref{#1})}

\newcommand\fig[1]{Fig.\,{\ref{#1}}}
\newcommand\sect[1]{Sect.\,{\ref{#1}}}

\newcommand\tab[1]{Table~\ref{#1}}

\newcommand{\beq}{\begin{equation}}
\newcommand{\eeq}{\end{equation}}
\newcommand{\bea}{\begin{eqnarray}}
\newcommand{\eea}{\end{eqnarray}}
\newcommand{\hf} {\frac{1}{2}}

\newcommand{\nn}{\nonumber\\}

\def\mr#1{{\mathrm{#1}}}

\def\tu{{\tilde u}}

\def\t{\tilde}

\begin{document}
\title{Asymptotic safety in the sine-Gordon model}

\author{J. Kov\'acs}
\affiliation{Department of Theoretical Physics, University of Debrecen,
P.O. Box 5, H-4010 Debrecen, Hungary}

\author{S. Nagy}
\affiliation{Department of Theoretical Physics, University of Debrecen,
P.O. Box 5, H-4010 Debrecen, Hungary}

\author{K. Sailer}
\affiliation{Department of Theoretical Physics, University of Debrecen,
P.O. Box 5, H-4010 Debrecen, Hungary}

\date{\today}

\begin{abstract}
In the framework of the functional renormalization group method
it is shown that the phase structure of the 2-dimensional sine-Gordon model possesses a
nontrivial UV fixed point which makes the model asymptotically safe.
The fixed point exhibits strong singularity similarly to the scaling found
in the vicinity of the infrared fixed point. The singularity signals the
upper energy-scale limit to the validity of the model. We argue that the sine-Gordon model
with a momentum-dependent wavefunction renormalization is in a dual connection with the massive sine-Gordon model.
\end{abstract}

\maketitle

\section{Introduction}

The 2-dimensional (2D) sine-Gordon (SG) model in Euclidean spacetime is one of the most
important scalar models \cite{Coleman:1974bu}. It has significant relevance starting from low energy condensed matter systems
\cite{2007PhRvL..99t7002B}, where it serves as an effective theory up to supersymmetry
\cite{Bajnok:2002vp,Bajnok:2003dk}, or string theory \cite{Hofman:2006xt}.
The model is integrable \cite{Zamolodchikov:1995xk,Delfino:1997ya,Mussardo:2004rw,Codello:2013iqa}
therefore it is possible to get analytic information from its physical properties.

The functional renormalization group (RG) treatment
\cite{Wetterich:1992yh,Berges:2000ew,Polonyi:2001se,Pawlowski:2005xe,Gies:2006wv}
for the SG model is capable of
describing the phase structure and identifying the Kosterlitz-Thouless (KT)
fixed point \cite{Berezinskii:1971,Kosterlitz:1973xp} of the model even in the local potential approximation
(LPA) \cite{Nandori:1999vi,Nagy:2006pq,Pangon:2009wk}. The perturbative RG treatment could account for the KT-type
essential scaling of the correlation length \cite{Amit:1979ab}.
The functional RG treatment also can describe the essential scaling
in equivalent models \cite{Grater:1994qx,VonGersdorff:2000kp}.
It is possible to investigate the SG model directly in the framework of the functional RG method
by taking into account the evolution of the wavefunction renormalization
\cite{Nagy:2009pj,Nagy:2010mf,Nandori:2011ss}, where the scaling of the coupling can be traced
from the perturbative regime to the infrared (IR) one. It was shown that there exists an
IR fixed point \cite{Nagy:2009pj,Nagy:2010mf}, which appears in the low energy scaling regime of the
broken symmetric phase. There are further models where the IR fixed point was identified successfully
\cite{Tetradis:1992qt,Kaplan:2009kr,Braun:2010tt,Nagy:2010mf,Nagy:2012np,Nagy:2012qz,Nagy:2012ef}.
Besides the existing fixed points it is questionable how the model behaves in the UV region of
the symmetric phase. The UV scaling of the coupling is irrelevant there, making it perturbatively
non-renormalizable. We argue that there exists a non-Gaussian UV fixed point (NGFP) in the SG model.
Our goal in this article is to clarify the UV scaling behavior of the model.
The UV NGFP makes the coupling finite, safe from divergences.
This is the main idea of the asymptotic safety \cite{Percacci:2007sz,Reuter:2007rv,Nagy:2012ef,Litim:2014uca}.
Nowadays the asymptotically safe models are in the focus in the RG treatment, since they
usually cannot be handled by perturbative tools. Recently the quantum Einstein gravity is
one of the most popular model which is investigated in the framework of the RG method
\cite{Reuter:1996cp,Reuter:2001ag,Reuter:2012id} and seems to show a UV NGFP making the model asymptotically
safe, even in its extensions \cite{Eichhorn:2011pc,Rechenberger:2012pm,Eichhorn:2012va,Dona:2013qba,Eichhorn:2013ug}.

There are several extensions of the SG model, the simplest one contains a simple mass term
\cite{Nagy:2004ey,Nagy:2006ue,Nandori:2010ij,Nandori:2012tc}. The mass is a relevant coupling
in the theory and breaks the periodic symmetry. In the UV limit the mass dies out therefore
the model shows an essential scaling. However in the IR, where the mass becomes more and more
important the essential scaling is replaced by a scaling belonging to a second order
phase transition \cite{Nagy:2012qz}.

In this article we make a further generalization of the SG model by taking into account a momentum
dependent wavefunction renormalization. Its Taylor expansion in the momentum provides further couplings.
There are couplings which are irrelevant and can cause troubles in the UV. We study how these
couplings affect the asymptotic safety and how they are related to the massive SG (MSG) model.

The paper is organized as follows. In \sect{sec:sg} the investigated models and the RG method
are introduced. In \sect{sec:as} we discuss how the asymptotic safety appears in the SG model.
The extensions of the SG model are treated in \sect{sec:zsg} and in \sect{sec:msg}.
The idea of duality among SG models are sketched in \sect{sec:dual}.
Finally, in \sect{sec:sum} the conclusions are drawn up.

\section{The sine-Gordon model}\label{sec:sg}

The functional renormalization group equation for the effective action of an 
Euclidean field theory is \cite{Wetterich:1992yh}
\beq\label{feveq}
\dot \Gamma_k=\hf\mr{Tr}\frac{\dot R_k}{R_k+\Gamma''_k}
\eeq
where the notations $^\prime=\partial/\partial\varphi$, $\dot{}=k\partial_k$ are used and the trace Tr
signals the integration over all momenta. The IR regulator function $R_k$
stands for removing the UV and IR divergences if necessary.
We have solved \eqn{feveq} by using the following form of the effective action
\beq\label{eaans}
\Gamma_k = \int_x\left[\frac{Z_k(\partial^2)}2 (\partial_\mu\phi)^2+V_k(\phi)\right],
\eeq
with the local potential
\beq
V_k(\phi) = u \cos(\phi).
\eeq
and the field-independent wave-function renormalization $Z_k(\partial^2)$ containing higher derivative terms.
The latter becomes momentum-dependent in Fourier-space and is parameterized as
\beq
Z_k = z+z_1 p^2,
\eeq
i.e. keeping only the first nontrivial term. The evolution equations for the couplings $u,z$ and $z_1$ are
\bea
\dot u &=& {\cal P}_1 \int_p k^2 {\cal D} \nn
\dot z &=& 2{\cal P}_0 \int_p k^2 V'''^2 {\cal D}^4\left\lbrack-\partial_{p^2}Z-2z_1p^2
+2p^2 (\partial_{p^2}Z)^2{\cal D}\right\rbrack\nn
{\dot z}_1&=&{\cal P}_0\int _p k^2 V'''^2{\cal D}^4\biggl\lbrack -2z_1\nn
&&~~~+\lbrack 24 p^2z_1 \partial_{p^2}Z+2(\partial_{p^2}Z)^2+12z_1^2p^4\rbrack{\cal D}\nn
&&~~~-\lbrack 12p^2(\partial_{p^2}Z)^3+36p^4z_1(\partial_{p^2}Z)^2\rbrack {\cal D}^2\nn
&&~~~+12p^4(\partial_{p^2}Z)^4{\cal D}^3\biggr\rbrack,
\eea
with the IR regulator $R_k=k^2$, the dressed propagator ${\cal D}=1/(Z_kp^2+k^2+V_k'')$ and
${\cal P}_n\ldots=(2\pi)^{-1}\int_0^{2\pi}d\phi \cos(n\phi)\ldots$ being the projection onto the field-independent subspace.

\section{Asymptotic safety}\label{sec:as}

\begin{center}
\begin{figure}[ht]
\epsfig{file=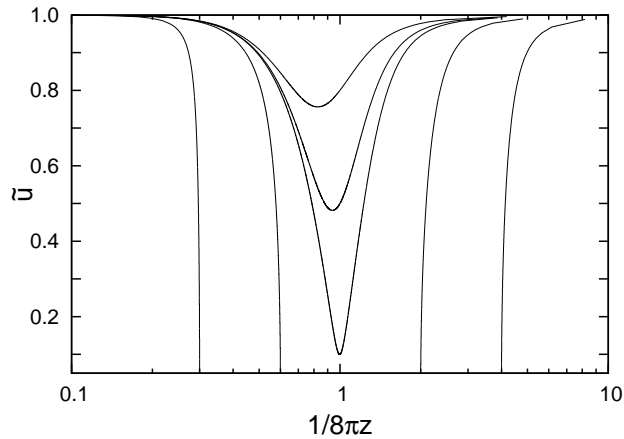,width=6cm,angle=-90}
\caption{\label{fig:sgphase}
The phase structure of the sine-Gordon model is presented. At $\t u=0$ we have a line of
fixed points. At $\t u=1$ and $z\to 0$ ($z\to\infty$) we have an infrared (ultraviolet)
non-Gaussian fixed point, respectively.
}
\end{figure}
\end{center}

First we restrict ourselves to consider the RG evolution of the dimensionless couplings $\t u = u/k^2$ and $z$ and
keep $\t z_1=0$. The evolution equations are \cite{Nagy:2009pj}
\bea
\label{sgRG}
\dot{\t u} &=& -2 \t u + \frac1{2\pi\t u z} \left[1-\sqrt{1-\t u^2}\right],\nn
\dot z &=& -\frac1{24\pi}\frac{\t u^2}{(1-\t u^2)^{3/2}}.
\eea
The phase space spanned by the couplings $\t u$ and $z$ are shown in \fig{fig:sgphase}.
If we try to look for the fixed points of the flow equations in \eqn{sgRG} then we cannot find any.
However if we Taylor expand the RG equations in the fundamental mode $\t u$,
then at $\t u^* = 0$ with arbitrary $z^*$ we have a line of fixed points. In certain
sense the line can be considered as the Gaussian fixed point(s) of the model. Its scaling
behavior shows that for $z^*>1/8\pi$ the section of the corresponding line is constituted by
UV attractive fixed points, while for $z^*<1/8\pi$ we have UV repulsive or IR attractive ones.
In other words in the vicinity of the lines of fixed points the evolution of the fundamental mode $\t u$
can be either relevant (for $1/z<8\pi$) or irrelevant (for $1/z>8\pi$).
The scaling regimes are separated by the KT point at $\t u^*_c = 0$ and $z^*_c=1/8\pi$.

The IR limit of the relevant scaling in the broken phase signals divergences and nonphysical
regimes. Similarly the UV limit of the irrelevant scaling can cause some trouble. Both
problems can be solved by finding non-trivial fixed points in these limits.

The IR fixed point was uncovered so far in \cite{Nagy:2009pj}.
It has been shown that the rescaling of the
original variables according to $\omega = \sqrt{1-\t u^2}$, $\chi=1/z\omega$
and $\partial_\tau = \omega^2 k\partial_k$ results in the evolution equations
\bea
\partial_\tau\omega &=& 2\omega(1-\omega^2)-\frac{\omega^2\chi}{2\pi}(1-\omega),\nn
\partial_\tau\chi &=& \chi^2\frac{1-\omega^2}{24 \pi}
-2\chi(1-\omega^2)+\frac{\omega\chi^2}{2\pi}(1-\omega).
\eea
Here the fixed point $\chi^*=0$ and $\omega^*=0$ can be identified
by the IR fixed point at $1/z^*=0$ and $\t u^*=1$, which is IR attractive.
Unfortunately one cannot find this fixed point from the original evolution equations in
\eqn{sgRG}
since it is situated at the singular point of the flow equation which makes the $\beta$-functions
divergent. It might raise the question whether the singularity is a numerical artifact due
to the truncations in the Fourier- and in the gradient expansion or it really exists.
However in $d=3$ we showed that the IR fixed point exists asymptotically if
we do not use any truncation \cite{Nagy:2012ef}. In the IR fixed point the soft modes play
crucial role in the spontaneously broken phase and seems to bring the quantum system to a classical one.
The huge amount of soft modes signals that the original degrees of freedom are not suitable to describe
the low energy behavior of the SG model.

\begin{center}
\begin{figure}[ht]
\epsfig{file=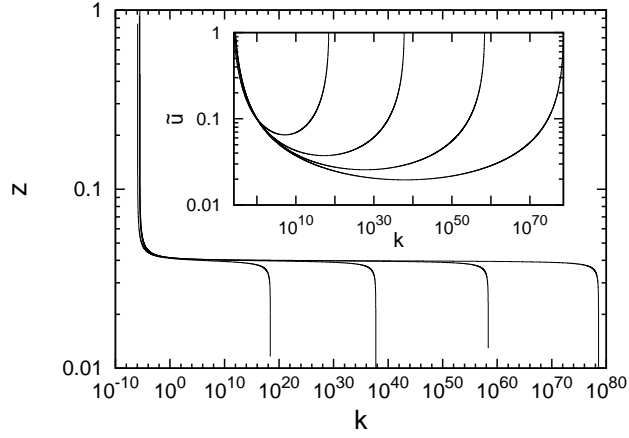,width=6cm,angle=-90}
\caption{\label{fig:uvz}
The evolution of the couplings $\t u$ and $z$ is shown. In the IR and the UV limits
the flows abruptly grow up or fall down at the same scale $k$.
}
\end{figure}
\end{center}

The problem of the irrelevant scaling in the symmetric phase can lead to divergences, which
can be avoided by finding a new non-trivial fixed point in the UV region. This is the
main idea of asymptotic safety. The UV NGFP keeps the value of the couplings and thus the
physical quantities finite. From the phase space in \fig{fig:sgphase} it is obvious that
the fundamental mode $\t u$ does not diverge but it tends to 1 as $z$ tends to 0 in the
UV limit. This motivated the choice of the logarithmic scaling in $z$.
Unfortunately the limit $\t u\to 1$ makes the RG equations in \eqn{sgRG} singular
as in the IR limit. Thus, we should approach the UV region of the SG model in a similar manner,
i.e. by a redefinition of the couplings. In the UV limit
the rescalings $\omega = \sqrt{1-\t u^2}$, $\zeta=z\omega$
and $\partial_\tau = z\omega^2 k\partial_k$ results in the evolution equations
\bea\label{sgRGuv}
\partial_\tau\omega &=& 2\zeta\omega(1-\omega^2)-\frac{\omega^2}{2\pi}(1-\omega),\nn
\partial_\tau\zeta &=& \left(2\zeta^2-\frac{\zeta}{24 \pi}\right)\left(1-\omega^2\right)
-\frac{\omega\zeta}{2\pi}(1-\omega).
\eea
Although the UV NGFP cannot be seen directly analytically from \eqn{sgRG},
the transformed equations in \eqn{sgRGuv} contain the corresponding fixed point at $\omega^*=0$ and $\zeta^*=0$.
This result can be affirmed by the fact that the ratio $du/dz$ calculated from the RG equations in \eqn{sgRG} 
tends to zero if $z\to 0$ and $\t u\to 1$.
In \fig{fig:uvz} the evolution of the couplings $\t u$ and $z$ is shown. In the IR limit
both couplings diverge at a certain value of $k_c$, which defines the IR correlation length.
In the UV limit the coupling $\t u$ tends to 1 making the evolution equations in \eqn{sgRG}
singular.
In \fig{fig:uvz} it is also shown that in the UV limit the coupling $z$ tends to zero, therefore
the kinetic term in the action freezes out, and only the potential term remains.
Similar situation appears in the confinement mechanism.
Although the couplings change abruptly at $k_c$ the trajectories do not show up singular behavior, since
the sudden increase of $\t u$ and the sudden decrease of $z$ compensate each other giving
regular flows. We obtained that in the vicinity of the UV NGFP the numerically calculated scaling gives $z=(1-\t u)^{3/2}$. 
The singular behavior in the UV limit defines a
corresponding correlation length $\xi=1/k_c$. The reduced temperature can be identified as the deviation
of the initial value of $z(\Lambda)$ to its critical value, i.e. $t\sim z(\Lambda)^*-z(\Lambda)$.
For the KT type phase transition we have
\beq\label{nukt}
\log\xi \propto t^{-\nu}.
\eeq
We determined the stopping scales $k_c$ for different initial couplings $z(\Lambda)$
and the numerical values gave $\nu\approx 0.51$ for the SG model.
It implies that the essential scaling can be uncovered in the UV limit, too, although it
originates from the slowing down of the flows around the KT point.

The point $z=0$, $\t{u}=1$ of the phase
space corresponds to the UV NGFP of the 2D SG model and it also makes the RG equations
singular. The UV singularity seems to mark the upper limit of the applicability of the SG model.
At higher energies the model needs new elementary excitations. Thinking
in the framework of the 2D XY model the excitations are represented by vortices which are made up by
concentric forms of spins \cite{Huang:1990via}. The blocking towards the increasing value of the scale $k$ results in
smaller and smaller vortices. At a certain scale $k_c$ the vortex reduces to a single spin.
This scale can be identified by the scale in the UV limit where the $\beta$-functions become singular.
At this scale the new elementary excitations should
be the single spins instead of the original vortices implying that in the UV limit the original
degrees of freedom of charged vortices should be replaced by a neutral spin system.

The functional RG method provides us both the high
and low energy scale limits of the applicability of the SG model. The low energy limit is usually indicated
by the IR fixed point in the broken phase. There the appearing condensate signals
the necessity of the new degrees of freedom. Anyway every model should also have an upper UV limit of its validity, since
presumably new degrees of freedom should appear at higher and higher energies.

\section{The sine-Gordon model with an irrelevant coupling}\label{sec:zsg}

\begin{center}
\begin{figure}[ht]
\epsfig{file=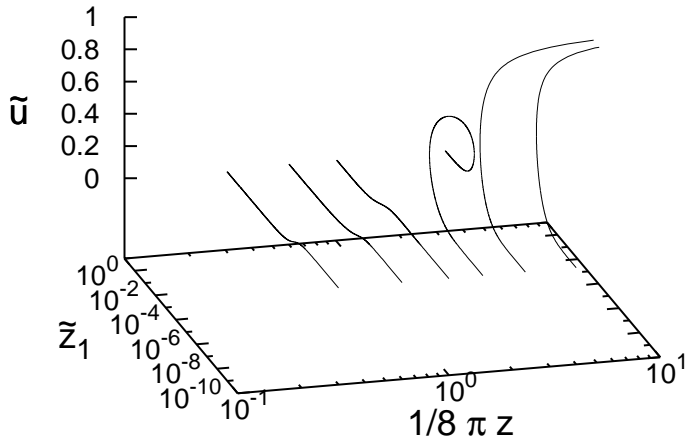,width=8cm,angle=-90}
\caption{\label{fig:z1phase}
The UV tendencies are plotted for the ZSG model with the choice $\t z_1(\Lambda) = 0.0$.
The coupling $\t z_1$ tends to infinity except when the singularity prevents it.
}
\end{figure}
\end{center}

Let us include the running of the coupling $\t z_1$ in the SG model, we abbreviate it as the ZSG model.
Then the RG equations for the dimensionless couplings become
\bea\label{z1sgRG}
\dot{\t u}&=&-2\t u-\frac{1}{\t u} \int_y  \bigg[1-\frac{\t Z y+1}{[(\t Z y+1)^2-\tu^2]^{1/2}} \bigg]\nn
\dot z&=&\frac{\t u^2}{4}\int_y \bigg[\frac{-(2\partial_y \t Z+4\t z_1 y)(\t Z y+1)}{[(\t Z y+1)^2-\t u^2]^{5/2}}
+\frac{y (\partial_y \t Z)^2(4(\t Z y+1)^2+\t u^2)}{[(\t Z y+1)^2-\t u^2]^{7/2}}\bigg]\nn
\dot{\t z}_1&=&2\t z_1+\frac1{48}\int_y \bigg[\frac{-24\t z_1(\t Z y + 1)}{[(\t Z y + 1)^2-\t u^2]^{5/2}}\nn
&&+\frac{(72\t z_1(\partial_y\t Z) y+6(\partial_y\t Z)^2+36\t z_1^2 y^2)(4(1+z y +\t z_1 y^2)^2+\t u^2)}
{[(\t Z y + 1)^2-\t u^2]^{7/2}} \nn
&&+\frac{(-36(\partial_y\t Z)^3 y-108z_1(\partial_y\t Z)^2 y^2)(\t Z y + 1)
(4(\t Z y + 1)^2+3\t u^2)}{[(\t Z y + 1)^2-\t u^2]^{9/2}}\nn
&&+\frac{(18(\partial_y\t Z)^4 y^2)(8(\t Z y + 1)^4+12(\t Z y+1)^2\t u^2+\t u^4)}
{[(\t Z y+1)^2-\t u^2]^{11/2}} \bigg],
\eea
with $\t Z = z y+\t z_1 y^2$. The momentum integral over $y=p^2$ cannot be performed analytically.
Although the $\beta$-function for the coupling $\t z_1$ is quite involved the numerical results show that the term coming
from the  inverse mass-dimension of $\t z_1$ drives the RG flow according to
\beq
\t z_1 \sim k^2,
\eeq
giving an irrelevant scaling in the IR limit. It implies that in the IR limit the inclusion of $\t z_1$ does not affect the
scaling of the SG model. There the phase transition remains a KT-type transition and we have the same
phase structure as in \fig{fig:sgphase}. However in the opposite scaling direction towards the UV limit $\t z_1$ diverges
and it removes the UV NGFP. The phase space of the ZSG model is sketched in \fig{fig:z1phase} for UV flows.
In the \eqn{z1sgRG} the RG equations become singular if the denominator
\beq\label{denom}
(\t Z y+ 1)^2-\t u^2= 0
\eeq
for any momentum. It can be zero if $\t u \to 1$. However the coupling $\t z_1$ grows up fast
which can prevent the singularity. The competition of these two scalings creates such
trajectories which constitute the two phases of the model.
In \fig{fig:z1phase} the UV symmetric phase contains such trajectories where $\t z_1\to \infty$.
The others will stop at finite $k$ and belong to the broken phase.
The UV correlation length can be defined similarly to the IR one, i.e., via
the stopping scale of $k_c$ where the UV evolution stops.
Its scaling shows that instead of the essential scaling there occurs a second order
Ising type phase transition according to
\beq\label{isingnu}
\xi\sim t^{-\nu},
\eeq
with the exponent $\nu=1/4$.
If $\t z_1(\Lambda)$ is increased then the broken phase starts to shrink.
The large initial value of $\t z_1$ can even overwrite the scaling behavior of $\t u$ in the way that the
original starting relevant scaling for $1/z>8\pi$ will be irrelevant and keeps the
trajectories in the UV symmetric phase. It is worth noting that a starting UV relevant scaling
of $\t u$ can turn to irrelevant (the middling trajectory in \fig{fig:z1phase}). Such a flow
did not appear in the original SG model.

The inclusion of the new coupling $\t z_1$ does not give a new phase. One can check it with the
sensitivity matrix \cite{Polonyi:2001se,Nagy:2006ue} which characterizes the deviation of the
running coupling w.r.t. the bare ones. According to \fig{fig:z1phase}
certain trajectories go to infinity in the UV and there are ones which stop at a finite scale $k$.
They can be neighboring trajectories which are adjacent in the IR limit and
then build up infinitely large distances from each other in the UV limit
signalling distinct phases according to the sensitivity matrix.
There are no further trajectories which can develop further singularities, implying that
we have no additional phases in the ZSG model.

\section{The sine-Gordon model with a relevant coupling}\label{sec:msg}

If we include a mass term in the potential
\beq
V_k(\phi) = \hf \t m^2\phi^2 + \t u \cos(\phi),
\eeq
then we have modified the SG model which looses its periodic symmetry. We call it the massive SG (MSG) model.
Neglecting the evolution of $\t z_1$ the RG equations are \cite{Nagy:2006ue,Nandori:2010ij}
\bea
\label{msgRG}
\dot{\t u} &=& -2 \t u + \frac1{2\pi\t u z} \left[1+\t m^2-\sqrt{(1+\t m^2)^2-\tu^2_1}\right],\nn
\dot z &=& -\frac1{24\pi}\frac{\tu_1^2}{((1+\t m^2)^2-\tu_1^2)^{3/2}},\nn
\dot{\t m}^2 &=& -2\t m^2.
\eea
The mass $\t m^2$ decouples from the other couplings and it scales according to the mass dimension of $m^2$,
\beq
\t m^2\sim k^{-2},
\eeq
so it is a relevant coupling as opposed to the previously introduced coupling $\t z_1$.
In the IR limit the MSG model shows a second order phase transition. There
we got $\nu=1/2$ \cite{Nagy:2012qz} for the exponent of the correlation length in \eqn{isingnu}.
In the IR limit the IR fixed point vanishes, since
the mass scales in relevant manner there. In the UV limit, the mass dies out due to its relevant scaling, therefore
the MSG and the SG models coincide, therefore it shows a KT-type infinite order phase transition.

\section{Duality}\label{sec:dual}

If we consider the phase diagram of the SG model with the couplings $\t u$ and $z$ in \fig{fig:sgphase}
then we can realize that the UV and the IR limits take place in a quite symmetric manner there.
If we make the following changes
\bea\label{dual}
k &\to& \frac1{k}\nn
z &\to& \frac1{z}
\eea
in the flow direction of the evolution (this is the way how the
UV limits are usually investigated) and the coupling then we obtain a qualitatively
similar phase structure. Similar relations were treated in \cite{Huang:1990via}. In this sense
the SG model shows a self duality. The advantage of this connection is that it is enough to investigate
a certain limitation of the model, and we can conclude the properties of the other limit from the former results.
The UV and the IR fixed points create a dual pair, and the lines of fixed points separated by
the KT point turn from one to the other under the dual transformation in \eqn{dual}. The KT fixed point
remains unchanged.

The duality can be extended to the ZSG and to the MSG models, too. In the former model
the original SG model contains a further irrelevant coupling while the MSG model has an additional relevant
one. We collected the appearing types of phase transitions in these models in \tab{tab:models}.
\begin{table}
\begin{center}
\begin{tabular}{|c||c|c|}
\hline
{\bf model} & {\bf UV} & {\bf IR} \\
\hline
\hline
SG & KT type, $\nu=1/2$ & KT type, $\nu=1/2$ \\
\hline
MSG & KT type, $\nu=1/2$ & Ising type, $\nu=1/2$ \\
\hline
z$_1$SG & Ising type, $\nu=1/4$ & KT type, $\nu=1/2$ \\
\hline
\end{tabular}
\end{center}
\caption{\label{tab:models} Summary of the SG-type models and their fixed points.}
\end{table}
In the ZSG model the IR scalings are not affected by an irrelevant couplings. The same is true
with opposite flow directions for the MSG model, there the UV scaling remains unchanged due to
the new relevant coupling. It implies that the IR ZSG and the UV MSG models behave as the
original SG model, they show an essential scaling.
However, in the other directions we have significant changes. In the UV limit
of the ZSG model and in the IR limit of the MSG model we can identify second order phase transitions.
It means that by completing the dual transformations in \eqn{dual} with
\beq
\t z_1 \to \t m^2,
\eeq
then the ZSG and the MSG models become a dual pair. It is not obvious that there exists any
relations between these models. The MSG model is the bosonized version of the 2D Quantum
Electrodynamics (or Schwinger model) \cite{Nagy:2004ey,Nagy:2006ue,Nandori:2010ij,Marian:2013zza},
and has no direct relation to the original SG model, since it is
well-known that after the bosonization it becomes the massive Thirring model \cite{Coleman:1974bu}.
The ZSG model can be related to the latter one, since there the wavefunction renormalization
is treated in a more precise way.

\section{Summary}\label{sec:sum}

The 2D sine-Gordon (SG) model and its extensions have been investigated by the functional renormalization group method
taking care of the deep infrared (IR) and the far ultraviolet (UV) limits.
We found that the SG model is asymptotically safe, i.e. it possesses a non-Gaussian fixed point (NGFP) in the UV limit.
The UV NGFP is hidden by the singularity which nicely traces out the upper limitations of the model.
The singularity has been identified in the IR limit, too. The similar behavior raises up the question
of the self-duality of the SG model for the UV and the IR limits.

We extended the SG model by taking into account the evolution of a further coupling in the momentum
dependent wavefunction renormalization. We showed that the obtained ZSG model has no UV NGFP and the UV
critical behavior is an Ising type phase transition. It reminds us to the massive SG (MSG) model,
where the IR limits showed a second order phase transition. This fact uncovered the dual relation between the ZSG and the MSG
models.

\section*{Acknowledgments}

This research was supported by the European Union and the State of Hungary, co-financed by the
European Social Fund in the framework of T\'AMOP 4.2.4.A/2-11-1-2012-0001 ‘National Excellence Program’.

\bibliography{nagy}

\end{document}